\documentstyle[aps,psfig,floats,epsf,twocolumn]{revtex}

\begin{document}
\tighten
\draft
\title{Collective Spin Fluctuation Mode and Raman Scattering in Superconducting Cuprates}

\author{F. Venturini$^1$, U. Michelucci$^2$, T.P. Devereaux$^3$, and
A.P. Kampf$^2$}
\address{$^1$Walther Meissner Institut, Bayerische Akademie der
  Wissenschaften, 85748 Garching, Germany}
\address{$^2$Theoretische Physik III, Elektronische Korrelationen und
Magnetismus, \\Institut f\"ur Physik, Universit\"at Augsburg, 86135 Augsburg, Germany}
\address{$^3$Department of Physics, University of Waterloo, Waterloo, ON,
Canada, N2L 3G1}
\address{~
\parbox{14cm}{\rm 
\medskip
Although the low frequency electronic Raman response in the
superconducting state of the cuprates can be largely understood in
terms of a d-wave energy gap, a long standing problem has been an
explanation for the spectra observed in the $A_{1g}$ polarization
orientations. We present calculations which suggest that the
peak position of the observed $A_{1g}$ spectra is due to a collective spin
fluctuation mode. 
\vskip0.05cm\medskip PACS numbers: 74.72.-h, 78.30.-j 71.10.-w}}

\maketitle

Electronic Raman scattering has proven to be a useful tool in exploring the 
superconducting state in the cuprate materials. The possibility 
of probing selectively electronic excitations in different regions of the 
Brillouin zone (BZ) by the choice of polarization geometries has allowed to 
explore the superconducting gap anisotropy. The successful explanation
of the Raman data in $B_{1g}$ and $B_{2g}$ scattering geometries
\cite{pol,tpd3} has provided one piece of evidence
for the by now widely accepted $d_{x^2-y^2}$ pairing symmetry in hole-doped 
cuprate superconductors \cite{Annett}. In the context of impurity
effects as a testing  
ground for unconventional superconductivity, the observed $\omega^3$
to $\omega$ crossover in the low frequency $B_{1g}$ Raman response
fits consistently with the power law crossovers at low temperatures in
the NMR relaxation rate and in  the magnetic penetration depth. An
even quantitatively consistent picture of electronic Raman scattering
and infrared conductivity was achieved when the  T-matrix
approximation in the ``dirty'' d-wave scenario is extended to include
a spatial extension of the impurity potential \cite{tpd1}.

However up to now the discrepancy between Raman data in $A_{1g}$ and
$B_{1g}, B_{2g}$ geometries has remained unresolved \cite{multi,Chub}. 
Previous results for the A$_{1g}$ scattering geometry were found to be very 
sensitive to changes in the Raman vertex function $\gamma({\bf k})$ 
\cite{multi} making a comprehensive explanation difficult for the 
experimental data in different cuprate materials.

In this paper we calculate the Raman response of a $d_{x^2-y^2}$
superconductor including the contribution from a collective spin
fluctuation (SF) mode which is identified with the
$(\pi,\pi,\pi)$-resonance (in short $\pi$-resonance)  near $\omega_R
\approx 41$ meV observed by inelastic neutron scattering (INS) on
bilayer cuprates
\cite{fong,mook}. Our results suggest that the $A_{1g}$ peak position is 
largely controlled by the strength and frequency of the $\pi$-resonance mode
which on the other hand does not affect the Raman response in the $B_{1g}$ and 
$B_{2g}$ channels. The inclusion of the collective SF mode allows for a
simultaneous fit of the Raman data in all channels in optimally doped
materials. Furthermore we find that the inclusion of the SF term significantly 
reduces the sensitivity to the special choice of the underlying tight-binding 
band structure, i.e. the sensitivity to the choice of the Raman vertex in 
$A_{1g}$ symmetry, resolving the previously encountered problems in the 
symmetry analysis of the light scattering amplitude \cite{multi,tpd3}.  

On the basis of the observation of the collective SF
mode in Y-123 and Bi-2212, we consider a bilayer model represented by a
tight-binding band structure
$$
\epsilon_{\bf k} = -2 t (\cos(k_x)+\cos(k_y))+4t'\cos(k_x) \cos(k_y)
- t_\perp({\bf k}), 
$$
with an inter-plane hopping given by \cite{bulut}
\begin{equation}
t_\perp({\bf k}) = 2 t_\perp \cos (k_z) (\cos(k_x)-\cos(k_y))^2
\label{tperp}
\end{equation}
where $k_z$ is $0$ or $\pi$ for the bonding or anti-bonding bands of the
bilayer, respectively.

The spin susceptibility ($\chi_s$) is modeled by extending the weak coupling 
form of a BCS superconductor in a $d_{x^2-y^2}$ pairing state to include 
antiferromagnetic spin fluctuations by a RPA form with an effective interaction
$\bar U$; i.e. $\chi_s({\bf q},i\omega)=\chi^0({\bf q},i\omega)/(1-\bar U\chi^0
({\bf q},i\omega))$ where \cite{bulut} 
\begin{equation}
\chi^0({\bf q},i\omega)={1\over{\beta}}\hbox{Tr} \sum_{{\bf k},i\omega'} 
\hat G({\bf k},i\omega') \hat G({\bf k+q},i\omega'+i\omega).
\end{equation}
Tr denotes the trace and $\beta=T^{-1}$. 
$\hat G({\bf k},i\omega)$ is the BCS Green's function in Nambu space
\begin{equation}
\hat G({\bf k},i\omega)=
\frac{i\omega \hat\tau_0 + \xi_{\bf k} \hat\tau_3 + \Delta_{\bf k} 
\hat\tau_1} {(i\omega)^2 -\xi_{\bf k}^2 - \Delta_{\bf k}^2}
\label{bcsg}
\end{equation}
with $\hat\tau_i$ (i=1,2,3) the Pauli matrices, $\hat\tau_0$ the
$2\times 2$ unit matrix, $\xi_{\bf k}=\epsilon_{\bf k}-\mu$ and
$\Delta_{\bf k}=\Delta_0[\cos(k_x)-\cos(k_y)]/2$.
This form of the spin susceptibility  contains a strong magnetic
resonance peak at ${\bf q}={\bf Q}\equiv (\pi,\pi,\pi)$ which was proposed 
\cite{bulut} to explain the INS resonance at energies near 41 meV in
Y-123 \cite{mook} and Bi-2212 \cite{fong}.
Other forms for the spin susceptibility can be straightforwardly used
within our model. However, the results are mainly determined by the
collective mode at ${\bf Q}$. Therefore we take the bilayer
susceptibility for a representative calculation.

The intensity of scattered light $I(\omega)$ in Raman experiments is 
proportional to the imaginary part of the response function for the effective 
density operator
\begin{equation}
\tilde \rho_{\bf q}=\sum_{{\bf k},\sigma} \hat\gamma ({\bf k})
c^\dag_{\sigma,{\bf k+q}} c_{\sigma,{\bf k}}
\label{rho}
\end{equation}
in the long wavelength limit ${\bf q\rightarrow 0}$. Specifically 
\begin{eqnarray}
I(\Omega)&\propto &\left(1+n(\Omega)\right)\hbox{Im} \chi(\Omega+i0^+)
\nonumber \\ 
\chi(i\Omega)&=&\int_0^{1/T}{\rm d}\tau\, e^{-i\Omega\tau}\,\langle T_\tau [
{\tilde\rho}(\tau),{\tilde\rho}(0)]\rangle ,
\end{eqnarray}
with the Bose function $n(\omega)$ and the time ordering operator
$T_\tau$. 

The bare Raman vertex $\hat\gamma({\bf k})=\tau_{3}\gamma({\bf k})$ in 
different scattering geometries are
classified according to the elements of the $D^{4h}$ point group. 
For the limiting case of vanishingly small scattered ($\omega_S$)
and incident ($\omega_I$) photon energies, it can be
represented in the effective-mass approximation (EMA)
\begin{equation}
\gamma({\bf k}) = \sum_{\alpha, \beta} e_\alpha^I \frac{\partial^2
\epsilon_{\bf k}}{\partial k_\alpha \partial k_\beta} e_\beta^S
\label{effmass}
\end{equation}
where ${\bf e}^I$ and ${\bf e}^S$ are the unit vectors for in-plane 
polarizations (i.e. $\alpha,\beta\in \{x,y\}$) of the incoming and the 
scattered light, respectively. 
Using Eq. 
(\ref{effmass}) and the bilayer tight-binding dispersion
Eq. \ref{tperp} we obtain
\begin{eqnarray} 
\gamma_{\bf k}^{B_{1g}}&=&2t\gamma_{\bf k}^d\left(1+{4t_\perp\over t}\cos(k_z)
[\cos(k_x)+\cos(k_y)]\right) \\
\gamma^{A_{1g}}_{\bf k}&=&2 t\gamma_{\bf k}^s-2 t_\perp \cos(k_z)
(\cos(2 k_x)+ \cos(2 k_y))\nonumber \\
&&-4\cos(k_x) \cos(k_y)[t'+2 t_\perp \cos(k_z)] 
\end{eqnarray}
where $\gamma_{\bf k}^{d,s}=(\cos(k_x) \mp \cos(k_y))/2$.
However, the EMA has a questionable region of
validity for all Raman measurements\cite{multi} on the cuprates since the 
incoming photons have energy $\sim 2$eV, which is on the order
of the bandwidth and the inter-band excitations according to local
density calculations\cite{lda}. EMA based arguments in previous works
about relative Raman intensities for different channels  
are therefore questionable \cite{Chub}. We hence consider other
forms for the vertices as well which obey the proper symmetry transformations.
For the $A_{1g}$ geometry some symmetry compatible choices are
$\gamma({\bf k})=\cos(k_x)+
\cos(k_y)$ and $\gamma({\bf k})=\cos(k_x)\cos(k_y)$. These basis functions
assign weight to different regions of the BZ and this is the reason why  
previous results for the $A_{1g}$ response were particularly sensitive to
the specific choice of the bare Raman coupling vertex. 

\begin{figure}
\vspace{5mm}
\centerline{\psfig{file=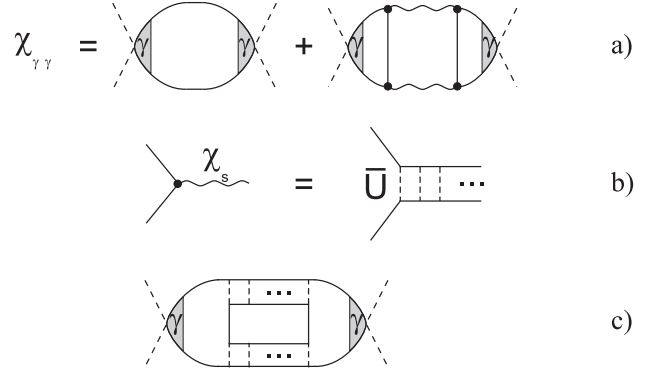,width=8.3cm,silent=}}
\vspace{.5cm}
\caption{a) Feynman diagrams for the Raman response function including 
pair-breaking quasiparticle excitations and ``2-magnon processes''. Dashed, 
wiggly, and solid lines represent photons, spin fluctuations, and fermionic 
propagators. $\gamma$ denotes the bare Raman vertex in a selected scattering 
geometry. b) Ladder series for the spin fluctuation propagator. c)
Explicit diagram for the spin fluctuation contribution.}
\label{diagr}
\end{figure}

The spin fluctuations lead to an additional contribution for the Raman response
via a 2-magnon like process as shown diagrammatically in Fig. \ref{diagr}
\cite{kampf}. Here, the SF propagator is incorporated in its RPA 
form for the bilayer (as described above) by the ladder diagram series with an 
effective on-site Hubbard interaction ${\bar U}$.

We therefore write the Raman response function at finite temperature as the sum
of a pair-breaking (PB) and a SF contribution
\begin{equation}
\chi_{\gamma\gamma}({\bf q},i\omega)=\chi^{PB}_{\gamma\gamma}({\bf q},i\omega)+
\chi^{SF}_{\gamma\gamma}({\bf q},i\omega) 
\end{equation}
with the Raman vertex 
specifying the scattering geometry. In the limit ${\bf q}
\rightarrow{\bf 0}$ the diagram for the SF contribution translates into
\begin{eqnarray}
\chi^{SF}_{\gamma\gamma}(i\Omega)&=&\displaystyle\frac{1}{\beta}\sum_{
{\bf q}',i\omega}\medskip V^\gamma({\bf q}',i\Omega,i\omega)\chi_s(-{\bf 
q'},-i\omega) \nonumber  \\
&&\times \chi_s({\bf q}',i\omega+i\Omega)V^\gamma({\bf q}',-i\Omega,-i\omega).
\label{chisf}
\end{eqnarray}
The vertex function $V^\gamma({\bf q}',i\Omega,i
\omega)$ includes the bare Raman vertex and is evaluated as 
\begin{eqnarray}
V^\gamma({\bf q}',i\Omega,i\omega)=\hbox{Tr}\bigg\{\displaystyle\frac{1}{\beta
}\sum_{{\bf k},i\omega'}\hat \gamma({\bf k})
\hat G({\bf k},i\omega'+i\Omega) \hat \tau_0 \bar U 
\nonumber\\
\hat G({\bf k+q}',i\omega'+i\Omega+i\omega) \tau_0 \bar U
\hat G({\bf k},i\omega')\bigg\} .
\label{vertex}
\end{eqnarray}
$i\Omega$, $i\omega$ denote bosonic and $i\omega'$ fermionic Matsubara 
frequencies.
Similarly, $\chi^{PB}_{\gamma\gamma}$ is evaluated as
\begin{equation}
\chi^{PB}_{\gamma\gamma}(i\Omega)={\rm{Tr}\over{\beta}}
\sum_{{\bf k},i\omega'}{\hat
\gamma}({\bf k})\hat G({\bf k},i\omega'){\hat\gamma}({\bf k})
\hat G({\bf k},i\omega'+i\Omega) .
\end{equation}

\begin{figure}
\centerline{%
\psfig{file=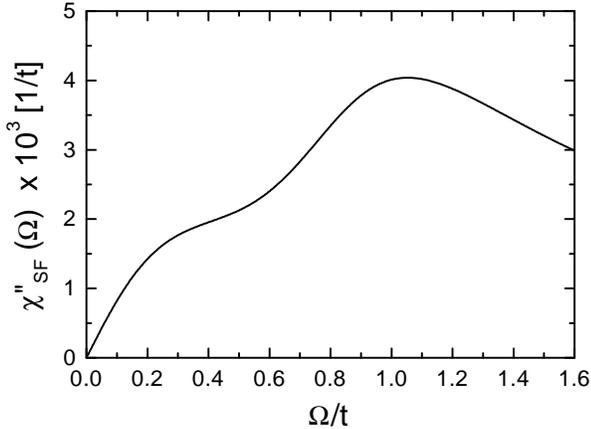,width=8.3cm,bbllx=63bp,bblly=125bp,bburx=607bp,bbury=529bp,clip=}}
\vspace{.4cm}
\caption{``2-magnon process'' contribution $\chi^{SF}_{\gamma\gamma}$ in 
the $A_{1g}$ geometry. The parameters used are defined in the text.}
\label{before}
\end{figure}

The total Raman response is calculated in the gauge invariant form which 
results from taking into account the long wavelength fluctuations of the order 
parameter to guarantee local charge conservation \cite{tpd3,note}. The
total Raman susceptibility thus follows as
\begin{equation}
\chi(i\Omega)=\chi_{\gamma \gamma}(i\Omega)-
\frac{\chi^2_{\gamma 1}(i\Omega)}{\chi_{11}(i\Omega)} 
\label{screening}
\end{equation}
where $\chi_{1\gamma}$ and $\chi_{11}$ are obtained by the replacement 
$\gamma({\bf k})\rightarrow 1$ in one or both bare Raman vertices in
the vertex function Eq. (\ref{vertex}). The analytical continuation to
the real axis is performed using Pad\'e approximants\cite{pade}.

The band structure parameters are chosen for all the numerical
calculations to be applicable to optimally doped systems:
$\langle n\rangle =0.85$, $t'/t=0.45$, $t_\perp/t=0.1$
\cite{multi} while the gap has been chosen as $\Delta_0/t=0.25$. 
We have evaluated the Raman response at a temperature $T/t=0.08$.

Let's first consider $\chi_{SF}$ alone.
In Fig. \ref{before} we plot $\chi''_{SF}$ for the $A_{1g}$ channel
versus frequency.
$\chi_{SF}$ is a convolution of two spin susceptibilities
(Eq. (\ref{chisf})) and 
is thus peaked at an energy near twice the magnetic resonance energy.
An important point is that in the $B_{1g}$ and $B_{2g}$
geometries the SF term introduces vanishingly small corrections to the
total response, rendering the presence of the 
SF term important only in the A$_{1g}$ geometry. This is due to the
sharpness in momentum space of the resonance peak at ${\bf Q}$ in the
SF propagator. In
fact, if the transfer is taken only at ${\bf Q}$, both the $B_{1g}$ and
$B_{2g}$ contributions to $\chi_{SF}$ vanish identically, as can been seen
from Eq. \ref{vertex}. Therefore for these channels the response is given
by the PB term alone.    

\begin{figure}
\centerline{%
\psfig{file=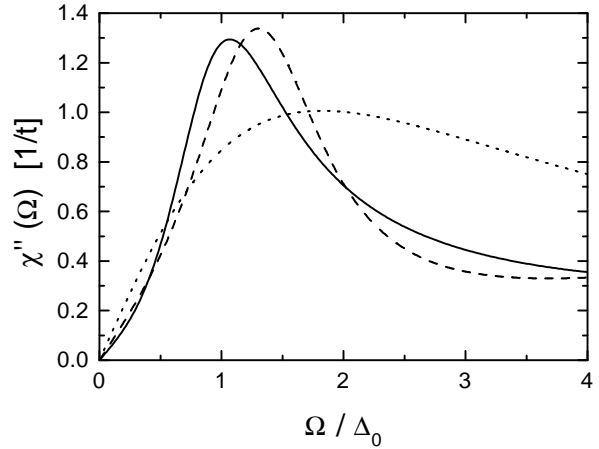,width=8.3cm,bbllx=63bp,bblly=120bp,bburx=607bp,bbury=529bp,clip=}}
\vspace{.4cm}
\caption{The total response in the $A_{1g}$ channel. Solid, dashed and
dotted lines correspond to $\bar U=0$, $\bar U/t=1.3$ and $\bar
U/t=2.0$ respectively.}
\label{after3}
\end{figure}

In Fig. \ref{after3} we illustrate
the frequency dependence of $\chi''(\Omega)$ in the
A$_{1g}$ geometry for three different values of the
effective interaction 
with both the PB and SF contributions included. 
The shape of the Raman response is modified varying $\bar U/t$ and in
particular the position of the resonance is shifted towards higher
energies for increasing $\bar U$.
With the inclusion of the SF term we now obtain a peak slightly above
$\Delta_0$ as observed experimentally in Y-123 and Bi-2212. 

A comment is in order on the relative magnitude of the SF and
the PB term. Comparing Figs. \ref{before} and
\ref{after3} it is clear that the SF term is much smaller than the PB term, 
but the effect of this new term is nevertheless visible since the backflow
(the second term in Eq. (\ref{screening})) mixes in a non trivial way
the two contributions. Since the
SF term varies as $\bar U^{4}$ in our model it starts to dominate for
larger $\bar U$, leading to
a shift in spectral weight out towards $2\omega_{R}$, which also changes
with $\bar U$.  
 
The experimental position of the peak is not sample dependent
in the $A_{1g}$ geometry as already mentioned,
and is almost the same in
different cuprates. On the other hand the theoretical
description with the PB term alone is very sensitive to Raman
vertex changes, which can produce variation of its position between
$\Delta_0$ and $2\Delta_0$ \cite{multi}, not allowing for a comprehensive
modeling of different cuprates.

\begin{figure}[t]
\centerline{%
\psfig{file=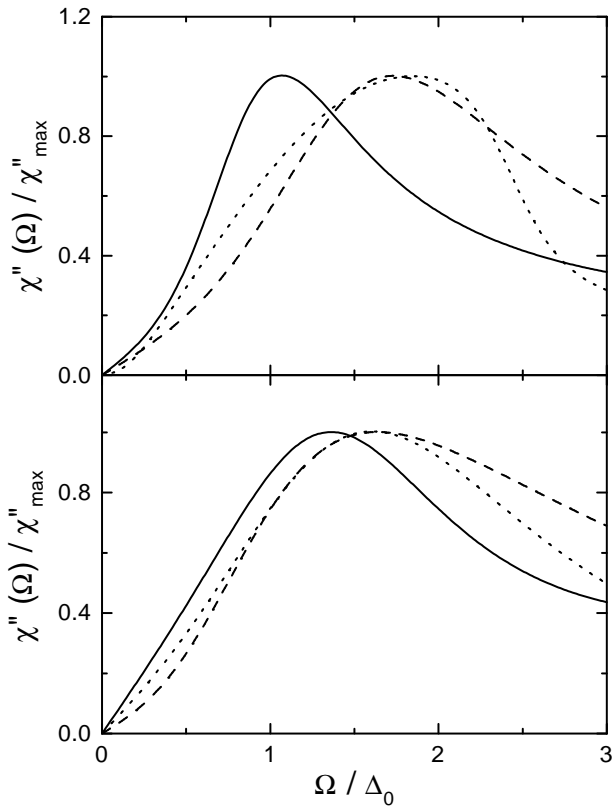,width=8.3cm,clip=,bbllx=15bp,bblly=50bp,bburx=541bp,bbury=747bp}}
\vspace{.4cm}
\caption{The total $A_{1g}$ response for different vertices: $\gamma$ obtained from
effective mass approximation (solid line), $\cos(k_x)+\cos(k_y)$
(dashed line), $\cos(k_x)\cos(k_y)$ (dotted line).}
\label{sens}
\end{figure}

In Fig. \ref{sens} we address the problem of the sensitivity of the
result to changes  
in the bare Raman vertex $\gamma({\bf k})$.
To investigate the effect of changes of the vertex
function, we have
calculated the final response using the three different forms for the
vertex $\gamma({\bf k})=\cos(k_x)+
\cos(k_y)$, $\gamma({\bf k})=\cos(k_x)\cos(k_y)$ and in the EMA
which posses the correct transformation properties required by
symmetry.
In the first panel of Fig. \ref{sens}, we have plotted the Raman response
for $\bar U=0$, i.e. the PB contribution alone, and in
the second panel for $\bar U/t=1.3$. 
All curves are renormalized to their peak height to allow for an
easier comparison.
Clearly the strong sensitivity to changes of the bare
Raman vertex (first panel) is much reduced when the SF term is added
(second panel).

\begin{figure}
\centerline{%
\psfig{file=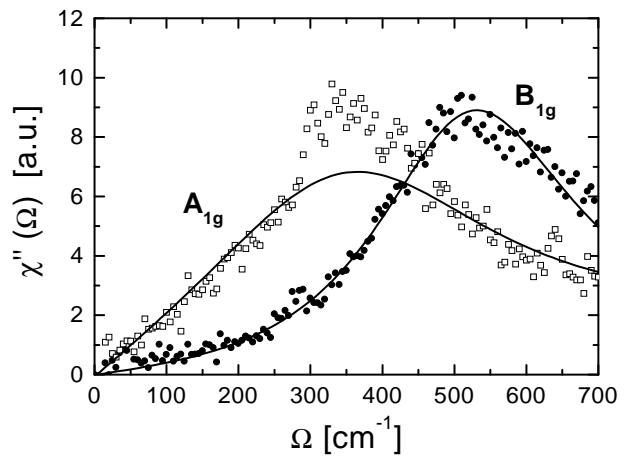,width=8.3cm,clip=,bbllx=73bp,bblly=127bp,bburx=600bp,bbury=520bp}}
\vspace{.4cm}
\caption{Comparison of the A$_{1g}$ and B$_{1g}$ response with Bi-2212
data. The parameters used are $t'/t=0.45$, $t=130$ meV,
$\Delta_0/t=0.25$, $\bar U/t=1.3$ and $\langle n \rangle = 0.85$. }
\label{fit}
\end{figure}

In Fig. \ref{fit} we compare the theoretical results with
experimental data on optimally doped 
Bi-2212 \cite{multi}. Adding the SF
contribution leads to a shift of the peak position from near $\sim
\Delta_0$ for $\bar U=0$ to higher
frequencies, and thus to a better agreement with the experimental
relative peak positions in $A_{1g}$ and $B_{1g}$ geometries.  
For the fit we have adjusted $t$ to achieve a good
agreement with the $B_{1g}$ channel, and then
adjusted $\bar U$ to match the $A_{1g}$ peak position. 

The value of $t$ obtained from the fit is
$t=130$ meV. This value has to be compared with $t \approx 105$ meV, which 
results from the condition $\omega_R \approx 40$ meV.
This slight discrepancy is most probably related to our simple modeling
of the propagators which neglects strong renormalizations from 
interactions as well as impurities.

From this work we conclude that including the SF contribution in the
Raman response solves the previously unexplained
sensitivity of the $A_{1g}$ response to small changes in the
Raman vertex.  
Also, within our model it is now possible to obtain the correct relative peak
positions of the $A_{1g}$ and the $B_{1g}$ scattering geometry.
Whereas the SF (two-magnon) contribution 
controls the $A_{1g}$ peak, the $B_{1g}$ and $B_{2g}$ scattering
geometries are essentially unaffected and determined by pair breaking
processes alone.

We would like to thank R. Hackl for numerous discussions.
One of the authors (F.V.) would like to thank the Gottlieb Daimler and 
Karl Benz Foundation for financial support.
This work was partially supported by the Deutsche Forschungsgemeinschaft
through SFB 484.


\begin{thebibliography}{99}
\bibitem{note}
Note that the term ``screening'' has been used to describe the backflow
yielding the second term in Eq. (\ref{screening}).
However, the backflow
is independent of the charge of the carriers and comes into play solely
from the pair interactions responsible for superconductivity. It yields
the well known Anderson-Bogoliubov mode which restores gauge invariance.
Details can be found in Ref. \cite{tpd3}.
\end{thebibliography}
\end{document}